\documentclass[11pt]{article}
\usepackage{amssymb,amsmath,latexsym}
\usepackage{mathrsfs}
\usepackage{tipa}
\usepackage{amsfonts}
\usepackage{graphicx}
\usepackage{subfigure}
\usepackage[title]{appendix}
\usepackage{color}
\usepackage{multirow}
\usepackage{amsbsy}
\usepackage{indentfirst}
\usepackage{longtable}
\usepackage{floatrow,authblk}
\usepackage{natbib}
\usepackage{enumitem}
\usepackage{scalerel}
\newcommand\wh[1]{\hstretch{2}{\widehat{\hstretch{.5}{#1}}}}
\usepackage{fancyhdr}   
\fancypagestyle{plain}{%
   \fancyhf{}
   \fancyfoot[C]{\tiny This paper is a preprint of a paper submitted to \it{IET Signal Processing.}}
   
}
\begin{document}
\title {Statistical inference for block sparsity of complex signals}
\author{Jianfeng Wang\thanks{Corresponding author : Jianfeng Wang\\
\hspace*{4.3 mm} Email: jianfeng.wang@umu.se}}
\author{Zhiyong Zhou}
\author{Jun Yu}
\affil{Department of Mathematics and Mathematical Statistics, Ume\aa{} University,\protect\\
SE 901 87 Ume\aa, Sweden}

\date{}
\maketitle
\abstract{Block sparsity is an important parameter in many algorithms to successfully recover block sparse signals under the framework of compressive sensing. However, it is often unknown and needs to be estimated. Recently there emerges a few research work about how to estimate block sparsity of real-valued signals, while there is, to the best of our knowledge, no investigation that has been conducted for complex-valued signals. In this paper, we propose a new method to estimate the block sparsity of complex-valued signal. Its statistical properties are obtained and verified by simulations. In addition, we demonstrate the importance of accurately estimating the block sparsity in signal recovery through a sensitivity analysis.}

\noindent

\section{Introduction}
Compressive sensing (CS) initially emerged around the year 2006 \citep{Donoho2006,Cands2006}. The aim of CS is to recover an unknown sparse signal $\mathbf{x}\in \mathbb{C}^N$ from $m$ noisy measurements $\mathbf{y} \in \mathbb{C}^m$:
\begin{align}
\mathbf{y}=A\mathbf{x}+\boldsymbol{\epsilon}
\end{align}
where $A\in \mathbb{C}^{m\times N}$ is a measurement matrix with $m\ll N$ satisfying certain incoherence condition, e.g. restricted isometry property \citep[Chapter 6]{Foucart2013}, and $\boldsymbol{\epsilon}\in\mathbb{C}^m$ is additive noise such that $\lVert\boldsymbol{\epsilon}\rVert_2\le \zeta$ for some $\zeta\ge0$.

There are several specific algorithms to recover $\mathbf{x}$ in $(1)$, e.g. orthogonal matching pursuit (OMP), compressive sampling matching pursuit (CoSaMP), iterative hard thresholding (IHT), and hard thresholding pursuit (HTP) \citep[Chapter 3]{Foucart2013}. All these algorithms require the value of sparsity as an input. Besides, in theory the minimal requirement on the number of measurements for a reliable recovery from $\mathbf{y}=A\mathbf{x}$ depends also on the sparsity level, e.g. $m\ge C\lVert \mathbf{x}\rVert_0\ln(N/\lVert \mathbf{x}\rVert_0)$ provided that $A$ is a subgaussian random matrix and $\mathbf{x}$ is a sparse signal, where $C$ is a constant and $\lVert \mathbf{x}\rVert_0=\sum_{i=1}^NI(|x_i|>0)$ is the sparsity of $\mathbf{x}$ \citep{Cands2008}. However, $\lVert \mathbf{x}\rVert_0$ is typically unknown in practice.

In addition to the simple sparse structure, a signal $\mathbf{x}$ can also possess another structure, i.e. blocks where the non-zero elements occur in clusters. A block signal $\mathbf{x}\in\mathbb{C}^N$ can be expressed as follows,
\begin{align}
\mathbf{x}=(\underbrace{x_1,\cdots, x_{d_1}}_{\mathbf{x}^{T}[1]},\underbrace{x_{d_1+1},\cdots, x_{d_1+d_2}}_{\mathbf{x}^{T}[2]},\cdots,\underbrace{x_{N-d_n+1},\cdots, x_N}_{\mathbf{x}^{T}[n]})^T, \label{signal}
\end{align}
where $N=\sum_{j=1}^n d_j$, $\mathbf{x}[j]$ is the $j$-th block of $\mathbf{x}$ over $\mathcal{I}=\{d_1,\cdots,d_n\}$ and $d_j$ is the length of the $j$-th block. Without loss of generality, throughout the paper we assume that $d_1=d_2=\cdots=d_n=d$, which implies that $N=nd$. By definition, the mixed $\ell_2/\ell_0$ norm $\lVert \mathbf{x}\rVert_{2,0}=\sum_{j=1}^{n}I(\lVert \mathbf{x}[j]\rVert_2>0)$ is the block sparsity of $\mathbf{x}$. {\color{black}Analogous to the recovery procedure for simple sparse signals, model (1) and corresponding algorithms can be used to recover a block sparse signal with some modifications, and the block sparsity of $\mathbf{x}$
plays also an important role both in the recovery algorithms and in determining the minimal required measurements number $m$.} It has been shown that using block information in CS can lead to a better signal recovery  \citep{Zamani2016}.
Thus, it is crucial to estimate the (block) sparsity beforehand in order to successfully recover (block) sparse signals. {\color{black} In fact, we can also express the block signal (2) in matrix form with dimension $n\times d$. Then instead of model (1), the multiple measurements vectors approach is capable to recover the matrix signals \citep{sun2009}. In the study, we investigate only the block sparsity measure of signals in vector form, and the measure can be easily generalized to signals in matrix form.}

Many signals are complex-valued in digital signal processing applications, e.g. medical imaging systems \citep{Graff2015}, digital communications systems \citep{Grami2013}, and radar systems \citep{Potter2010}. As pointed out by \citet{Sharif-Nassab2012}, there was lack of developments of sparsity estimation for complex-valued signals. It coincides with the fact that the inference procedures of sparsity in \citet{zy} and \citet{l2} are only valid for real-valued signals. {\color{black}Therefore the main purpose of the study is to introduce a method to estimate block sparsity of a complex-valued signal, which is one of the main contributions of the paper. The another contribution is that we substantiate the importance of accurately estimating the block sparsity in signal recovery through through a sensitivity analysis.}

The paper is organized as follows. In Section 2, we review the block sparsity measures in literature. Afterward, we explain the work flow of block sparsity estimation for complex-valued signals and discuss recovery algorithms that could be used to illustrate the importance of accurately estimating the block sparsity in Section 3. In Section 4, we introduce the measure of block sparsity and its statistical properties. The numerical justification is described in Section 5 and simulation results are presented in Section 6. Section 7 is devoted to the conclusion.

\section{Block sparsity measures}
Generally, there are two kinds of simple sparse signals that can be recovered by CS, i.e. (strictly) sparse signals and compressible signals \citep{Foucart2013,Cheng2015}. A signal is strictly sparse if most of its elements are zero. Compressible signal implies that the signal is not sparse but it can be well approximated by a sparse signal. Traditionally, $\ell_0$ norm, i.e. $\lVert \mathbf{x}\rVert_0$, has been used to measure the sparsity of a signal $\mathbf{x}=(x_1,x_2,\cdots,x_N)^T\in\mathbb{C}^N$. As mentioned in \citet{l1}, $\lVert \mathbf{x}\rVert_0$ was usually assumed to be a fixed and known value to recover the signal without taking its uncertainty into account, and there was no method developed to estimate it in literature, which has also been pointed out in \citet{Ward2009,eldar_2009}. However, a strictly sparse signal is rarely observed in practice, instead compressible signals are often received. Since the elements of such a signal around zero can often be ignored for reasons, e.g. they are introduced by noise, the $\ell_0$ norm is not appropriate to describe the sparsity of compressible signals. \citet{l1} proposed the following quantity  \[s(\mathbf{x})=\frac{\lVert \mathbf{x}\rVert_1^2}{\lVert \mathbf{x}\rVert_2^2},\] for measuring the sparsity of a compressible signal. However, given a signal $\mathbf{x}, s(\mathbf{x})$ is a fixed value which is not adjustable with varying noise levels. To overcome this drawback, Lopes (2016) generalized this measure by introducing \begin{align}
s_{\alpha}(\mathbf{x})=\left(\frac{\lVert \mathbf{x}\rVert_\alpha}{\lVert \mathbf{x}\rVert_1}\right)^{\frac{\alpha}{1-\alpha}},
\end{align}
for $\alpha\notin\{0,1,\infty\}$, where the $\ell_\alpha$ norm is defined as $\lVert \mathbf{x}\rVert_{\alpha}=\left(\sum_{i=1}^{N}|x_i|^{\alpha}\right)^{1/\alpha}$ for any $\alpha>0$. The cases for $\alpha\in\{0,1,\infty\}$ are evaluated by limits:
\begin{align}
s_0(\mathbf{x})&=\lim\limits_{\alpha\rightarrow 0}s_{\alpha}(\mathbf{x})=\lVert \mathbf{x}\rVert_0\\
s_1(\mathbf{x})&=\lim\limits_{\alpha\rightarrow 1}s_{\alpha}(\mathbf{x})=\exp(H_1(\pi(\mathbf{x})))\\
s_\infty(\mathbf{x})&=\lim\limits_{\alpha\rightarrow \infty}s_{\alpha}(\mathbf{x})=\frac{\lVert \mathbf{x}\rVert_1}{\lVert \mathbf{x}\rVert_{\infty}},
\end{align}
where $H_1(\pi(\mathbf{x}))=-\sum_{j=1}^N\pi_j(\mathbf{x})\ln\pi_j(\mathbf{x})$ is the ordinary Shannon entropy with $\pi(\mathbf{x})\in\mathbb{R}^N$ and its entries $\pi_j(\mathbf{x})=\frac{|x_j|}{\lVert \mathbf{x}\rVert_1} $. $s_\alpha(\mathbf{x})$ is a non-increasing function with respect to $\alpha$ and $s_\alpha(\mathbf{x}) \in [s_\infty(\mathbf{x}),s_0(\mathbf{x})]$. In other words, $\alpha$ determines the sparsity level of a compressible signal. For instance, if a compressible signal comprises larger noise, a larger $\alpha$ can be chosen to achieve a sparser signal that approximates the original noise free signal and vice versa. Lopes (2016) also provided statistical inference for $s_{\alpha}(\mathbf{x})$ with $\alpha\in(0,2]$ by random linear projections through the measurement matrix $A$ using independent and identically distributed (i.i.d.) univariate symmetric $\alpha$-stable random variables. {\color{black}Another application of the symmetric $\alpha$-stable distribution in CS can be found in \cite{Javaheri2018}, which proposed a continuous mixed $\ell_p$ norm for the sparse recovery. Interested readers are referred to the paper for details.}

 Similar to the classification for the simple sparse signals, a block sparse signal can be either (strictly) block sparse or block compressible. Like the definition for compressible signals, a block compressible signal can be well approximated by a block sparse signal. To quantify the block sparsity, \citet{zy} proposed a block sparsity measure, which extends the sparsity measure in (3). In the same manner as $\lVert \mathbf{x}\rVert_0$ improper to compressible signals, $\lVert \mathbf{x}\rVert_{2,0}$ is not proper to measure the sparsity of a block compressible signal. Hence, Zhou and Yu (2017) introduced an entropy based block sparsity measure:
\begin{align}
k_{\alpha}(\mathbf{x})=\left(\frac{\lVert \mathbf{x}\rVert_{2,\alpha}}{\lVert \mathbf{x}\rVert_{2,1}}\right)^{\frac{\alpha}{1-\alpha}}
\end{align}
for $\alpha\notin\{0,1,\infty\}$, where the mixed $\ell_2/\ell_\alpha$ norm is defined as $\lVert \mathbf{x}\rVert_{2,\alpha}=\left(\sum_{j=1}^{n}\lVert \mathbf{x}[j]\rVert_2^{\alpha}\right)^{1/\alpha}$ for any $\alpha>0$. The cases of $\alpha\in\{0,1,\infty\}$ are evaluated by limits: \begin{align}
k_0(\mathbf{x})&=\lim\limits_{\alpha\rightarrow 0}k_{\alpha}(\mathbf{x})=\lVert \mathbf{x}\rVert_{2,0}\label{app0}\\ k_1(\mathbf{x})&=\lim\limits_{\alpha\rightarrow 1}k_{\alpha}(\mathbf{x})=\exp(H_1({\pi}(\mathbf{x})))\\ k_\infty(\mathbf{x})&=\lim\limits_{\alpha\rightarrow \infty}k_{\alpha}(\mathbf{x})=\frac{\lVert \mathbf{x}\rVert_{2,1}}{\lVert \mathbf{x}\rVert_{2,\infty}},
\end{align}
where ${\pi}(\mathbf{x})=\big(\frac{\lVert\mathbf{x}[1]\rVert_2}{\lVert\mathbf{x}\rVert_{2,1}}, \frac{\lVert\mathbf{x}[2]\rVert_2}{\lVert\mathbf{x}\rVert_{2,1}}, \cdots, \frac{\lVert\mathbf{x}[n]\rVert_2}{\lVert\mathbf{x}\rVert_{2,1}}\big)$ and $\lVert \mathbf{x}\rVert_{2,\infty}=\max\limits_{1\leq j\leq n}\lVert \mathbf{x}[j]\rVert_{2}$. Like $s_\alpha(\mathbf{x})$, $k_\alpha (x)$ is a non-increasing function with respect to $\alpha$ and $k_\alpha(\mathbf{x}) \in [k_\infty(\mathbf{x}),k_0(\mathbf{x})]$. It is easy to see that $s_{\alpha}(\mathbf{x})$ is a special case of $k_{\alpha}(\mathbf{x})$ with $d=1$. It is important to notice that in addition to that $k_{\alpha}(\mathbf{x})$ could be used to measure block sparsity of a block compressible signal by adjusting $\alpha$, it can also approximate block sparsity of a block sparse signal shown in (\ref{app0}). As the block sparsity measure, $k_{\alpha}(\mathbf{x})$, of $\mathbf{x}$ depends also on the block size $d$, in the following context, we use $k_{\alpha,d}(\mathbf{x})$ to denote the block sparsity measure of $\mathbf{x}$ with block size $d$.

\section{{\color{black}Problem formulation}}
 In this study, we introduce a method to estimate block sparsity of complex-valued signal by making a reversible transformation to the signal so that 1) the transformed signal keeps all the original information and 2) the transformed signal is real-valued. In this way, we can adopt the inference procedure of block sparsity estimation for real-valued signals in \citet{zy}. The adaptation is rather straightforward but it has not been touched in existing literature to the best of our knowledge.

 Let's first introduce notations for complex-valued signals. Let $\mathbf{x}=\mathbf{a}+\sqrt{-1}\mathbf{b}\in\mathbb{C}^N$ with the real part $\mathbf{a}\in\mathbb{R}^N$ and imaginary part $\mathbf{b}\in\mathbb{R}^N$. We denote the $i$-th components of the vector by $x_i, a_i, b_i$ with $x_i=a_i+\sqrt{-1}b_i$. With these notations, to estimate the sparsity of a complex-valued signal $\mathbf{x}$ (of block size $d=1$), instead of taking the absolute value, we transform $\mathbf{x}$ to a $2N$-length real-valued signal $\tilde{\mathbf{x}}$ with block size $d=2$, and the $i$-th block is the real and imaginary components of $x_i$ for $i\in\{1,2,\cdots,N\}$, i.e. \begin{align}
\tilde{\mathbf{x}}=(\underbrace{a_1,b_1}_{\tilde{\mathbf{x}}^{T}[1]},\underbrace{a_2,b_2}_{\tilde{\mathbf{x}}^{T}[2]},\cdots,\underbrace{a_N,b_N}_{\tilde{\mathbf{x}}^{T}[N]})^T.
\end{align}
It is obvious that this transformation is reversible and the transformed signal $\tilde{\mathbf{x}}$ keeps all the information from the original $\mathbf{x}$. It is worth noticing that the sparsity, $s_{\alpha}(\mathbf{x})$ or $k_{\alpha,1}(\mathbf{x})$, of the complex-valued signal $\mathbf{x}$ equals to the block sparsity, $k_{\alpha,2}(\tilde{\mathbf{x}})$, of the transformed real-valued signal $\tilde{\mathbf{x}}$. This approach can be easily generalized to the case when $d>1$, and it holds that the block sparsity $k_{\alpha,d}(\mathbf{x})$ equals to the block sparsity $k_{\alpha,2d}(\tilde{\mathbf{x}})$ with \begin{align}
\tilde{\mathbf{x}}=(\underbrace{a_1,b_1,\cdots, a_d,b_d}_{\tilde{\mathbf{x}}^{T}[1]},\underbrace{a_{d+1},b_{d+1},\cdots, a_{2d}, b_{2d}}_{\tilde{\mathbf{x}}^{T}[2]},\cdots,\underbrace{a_{N-d+1},b_{N-d+1},\cdots, a_N,b_N}_{\tilde{\mathbf{x}}^{T}[n]})^T.
\end{align}
In summary, we have $k_{\alpha,d}(\mathbf{x})=k_{\alpha,2d}(\tilde{\mathbf{x}})$ with $d\ge 1$, and $s_{\alpha}(\mathbf{x})=k_{\alpha,1}(\mathbf{x})$. The problem of estimating the block sparsity of a complex-valued signal $\mathbf{x}$ with block size $d$ is therefore transformed to that of estimating the block sparsity of a real-valued signal $\tilde{\mathbf{x}}$ with block size $2d$.

To make statistical inference on $k_{\alpha,2d}(\tilde{\mathbf{x}})$, we use multivariate isotropic symmetric $\alpha$-stable random projections, which was initially introduced in \citet{l2} and extended to real-valued block sparse signals in \citet{zy}. The asymptotic distribution of the estimator of $k_{\alpha,2d}(\tilde{\mathbf{x}})$ has the same form as in \citet{zy}.

{\color{black}
Given a block sparsity estimated from the measurements, a natural but very essential question is how important a good estimate of block sparsity is for signal recovery in practice? There are a variety of recovery algorithms available for sparse signals. Some of them do not need to specify the sparsity level in advance. For instance, OMP based algorithms in \cite{Azizipour2019,Do2008}, and  Bayesian model based algorithms in \citet{Cao2018, Meng2018, Korki2016, Zhang2016}.
However, they are not considered in the study, since the knowledge of sparsity level is also very useful to decide the minimal number of required measurements. Accordingly, with extension from the simple sparse signals, recovery algorithms can be generalized for block sparse signals, e.g. model based CoSaMP in \citet{Baraniuk2010, Baron2009}, Block OMP in \citet{Eldar2010}, and Group
Basis Pursuit in \citet{Eldar2009}.
In the study}, we select the model based CoSaMP algorithm described in \citet{Baraniuk2010} to investigate how the performance of block sparse signal recovery can be influenced by the block sparsity. {\color{black} The properties, e.g. convergence, of the algorithm can be found in \citet{Baraniuk2010}.}

\section{{\color{black} The proposed} method}

In this section, we introduce the statistical inference for the block sparsity of unknown real-valued signal $\tilde{\mathbf{x}}$ (transformed from the original complex-valued signal ${\mathbf{x}}$) through random linear projections by using multivariate isotropic and symmetric $\alpha$-stable random vectors. We first give the definition of the multivariate centered isotropic symmetric $\alpha$-stable distribution.\\

\noindent
{\bf Definition 1.} For $d\geq 1$, a $d$-dimensional random vector $\mathbf{v}$ has a centered isotropic symmetric $\alpha$-stable distribution if there exist constants $\gamma>0$ and $\alpha\in(0,2]$ such that its characteristic function has the form \begin{align}
E[\exp(\sqrt{-1}\mathbf{u}^{T}\mathbf{v})]=\exp(-\gamma^{\alpha}\lVert \mathbf{u}\rVert_2^\alpha),\,\,\,\text{for all $\mathbf{u}\in\mathbb{R}^d$}.
\end{align}
We denote the distribution by $\mathbf{v}\sim S(d,\alpha,\gamma)$, and $\gamma$ is a dispersion of the distribution.\\

\noindent
{\bf Remark 1.} The family of multivariate centered isotropic and symmetric $\alpha$-stable distributions covers two well-known members: One is multivariate spherical symmetric Cauchy distribution with zero mean and identity covariance matrix when $\alpha=1$ and $\gamma=1$ \citep{Press1972}. The other is standard multivariate normal distribution when $\alpha=2$ and $\gamma=\frac{\sqrt 2}{2}.$\bigskip

Next, we estimate the $\lVert\tilde{\mathbf{x}}\rVert_{2,\alpha}^{\alpha}$ with block size $2d$ by using the random linear projection measurements: \begin{align}
y_i=\langle \mathbf{r}_i,\tilde{\mathbf{x}}\rangle+\sigma\varepsilon_i, \,\,\,i=1,2,\cdots,m, \label{model}
\end{align}
where $\mathbf{r}_i\in\mathbb{R}^{2N}$ is i.i.d random vector, and $\mathbf{r}_i=(\mathbf{r}_{i1}^T,\cdots,\mathbf{r}_{in}^T)^T$ with $\mathbf{r}_{ij},j\in\{1,\cdots,n\}$ i.i.d drawn from $S(2d,\alpha,\gamma)$. $\sigma$ is a variance parameter, the noise term $\varepsilon_i$'s are assumed to be i.i.d from a distribution $F_0$ having its characteristic function $\varphi_0$, and the sets $\{\varepsilon_1,\cdots,\varepsilon_m\}$ and $\{\mathbf{r}_{1},\cdots,\mathbf{r}_{m}\}$ are independent. $F_0$ is assumed to be symmetric about $0$ for simplicity, and it has finite first moment but may have infinite variance. Here we introduce a lemma that is useful in the estimation procedure, which is Lemma 3 in \citet{zy}.\\

\noindent
{\bf Lemma 1.} Let  $\tilde{\mathbf{x}}=( \tilde{\mathbf{x}}^T[1],\tilde{\mathbf{x}}^T[2],\cdots,\tilde{\mathbf{x}}^T[n])^T \in \mathbb{R}^{2N}$ and $\mathbf{r}_i=(\mathbf{r}_{i1}^T,\cdots,\mathbf{r}_{in}^T)^T$ with $\mathbf{r}_{ij}, i\in\{1,\cdots,m\}$ and $j\in\{1,\cdots,n\},$ i.i.d drawn from $S(2d,\alpha,\gamma)$ with $\alpha\in (0,2]$ and $\gamma>0$, then every random variable $\langle \mathbf{r}_i,\tilde{\mathbf{x}}\rangle$ has the distribution $S(1,\alpha,\gamma\lVert{\tilde{\mathbf{x}}}\rVert_{2,\alpha})$.\\

Now we are ready to present the estimation procedure by using the characteristic function. We apply the model (\ref{model}) to estimate $\lVert\tilde{\mathbf{x}}\rVert_{2,1}$ and
$\lVert\tilde{\mathbf{x}}\rVert_{2,\alpha}^{\alpha}$ with sample size $m$ equals to $m_1$ and $m_{\alpha}$, respectively. We will only describe the procedure of how to estimate $\lVert\tilde{\mathbf{x}}\rVert_{2,\alpha}^{\alpha}$ for any $\alpha\in(0,2]$, since $\lVert\tilde{\mathbf{x}}\rVert_{2,1}$ is a special case of $\lVert\tilde{\mathbf{x}}\rVert_{2,\alpha}^{\alpha}$ with $\alpha=1$.
Through Definition 1 and Lemma 1, we have the characteristic function of $y_i$:
\begin{align}
&\Psi(t)=E[\exp(\sqrt{-1}ty_i)]=\exp(-\gamma^{\alpha}\lVert\tilde{\mathbf{x}}\rVert_{2,\alpha}^{\alpha}|t|^{\alpha})\cdot\varphi_0(\sigma t),
\end{align}
which implies that
\begin{align}
\lVert\tilde{\mathbf{x}}\rVert_{2,\alpha}^{\alpha}=-\frac{1}{\gamma^{\alpha}|t|^\alpha}\ln \left|\mathrm{Re}\left(\frac{\Psi(t)}{\varphi_0(\sigma t)}\right)\right|.
\end{align}
As an estimator for $\Psi(t)$, we use the empirical characteristic function: $$\widehat{\Psi}_{m_\alpha}(t)=\frac{1}{m_\alpha}\sum\limits_{i=1}^{m_\alpha}e^{\sqrt{-1}ty_i}.$$
Consequently, the estimator of $\lVert\tilde{\mathbf{x}}\rVert_{2,\alpha}^{\alpha}$ can be obtained by
\begin{align}
\wh{\lVert\tilde{\mathbf{x}}\rVert}^{\alpha}_{2,\alpha}=-\frac{1}{\gamma^{\alpha}|t|^\alpha}\ln \left|\mathrm{Re}\left(\frac{\widehat{\Psi}_{m_\alpha}(t)}{\varphi_0(\sigma t)}\right)\right|, \label{esti}
\end{align}
when $t\neq 0$ and $\varphi_0(\sigma t)\neq 0$. By this, the estimator of $k_{\alpha,2d}(\tilde{\mathbf{x}})$ can be obtained by using $\wh{\lVert\tilde{\mathbf{x}}\rVert}_{2,\alpha}^{\alpha}$ and $\wh{\lVert\tilde{\mathbf{x}}\rVert}_{2,1}$:
\begin{align}
\widehat{k}_{\alpha,2d}(\tilde{\mathbf{x}})=\frac{\left(\wh{\lVert\tilde{\mathbf{x}}\rVert}_{2,\alpha}^{\alpha}\right)^{\frac{1}{1-\alpha}}}
{\left(\wh{\lVert\tilde{\mathbf{x}}\rVert}_{2,1}\right)^{\frac{\alpha}{1-\alpha}}}.
\end{align}
\\
Since the estimator (\ref{esti}) is a function of $t$, we should find an suitable estimator for $t$. Adapting the Theorem 1 in \cite{zy} to complex-valued signals, an optimality criterion for the choice of $t$ is given in the following proposition. \\

\noindent
{\bf Proposition 1}. Let $\alpha\in (0,2]$, $\rho_\alpha=\sigma/(\gamma\lVert\tilde{\mathbf{x}}\rVert_{2,\alpha})$, and $\hat{t}$ be any function of $\{y_1,y_2,\cdots,y_{m_\alpha}\}$ that satisfies
\begin{align}
\gamma\,\hat{t}\,\lVert\tilde{\mathbf{x}}\rVert_{2,\alpha}\overset{P}\longrightarrow c_\alpha,
\end{align}
as $(m_\alpha, N)\to \infty$ for some constant $c_\alpha \neq 0$ and $\varphi_0(\rho_\alpha c_\alpha ) \neq 0$. Then the estimator $\wh{\lVert\tilde{\mathbf{x}}\rVert}_{2,\alpha}^{\alpha}$ satisfies
\begin{align}
\sqrt{m_\alpha}\left(\frac{\wh{\lVert\tilde{\mathbf{x}}\rVert}_{2,\alpha}^{\alpha}}{\lVert\tilde{\mathbf{x}}\rVert_{2,\alpha}^\alpha}-1\right)\overset{D}\longrightarrow N\Big(0,\theta_\alpha(c_\alpha,\rho_\alpha)\Big),
\end{align}
as $(m_\alpha, N)\to \infty$, where
\begin{align}
\theta_\alpha(c_\alpha,\rho_\alpha)=\frac{1}{|c_\alpha|^{2\alpha}}\left(\frac{\exp(2|c_\alpha|^{\alpha})}{2\varphi_0(\rho_\alpha
	|c_\alpha|)^2}+\frac{\varphi_0(2\rho_\alpha|c_\alpha|)}{2\varphi_0(\rho_\alpha|c_\alpha|)^2}\exp((2-2^\alpha)|c_\alpha|^\alpha)-1\right),
\end{align} and it is strictly positive.\\

The optimal estimator of $t$, $\hat{t}_{\mathrm{opt}}$, should minimize the limiting variance $\theta_\alpha(c_\alpha,\rho_\alpha)$. In this study, we adopt $\hat{t}_{\mathrm{pilot}}$ for simplicity, which is an intermediate result of computing $\hat{t}_{\mathrm{opt}}$, although it may not be optimal in terms of limiting variance. The pilot estimate of $t$ is given by $\hat{t}_{\mathrm{pilot}}=\min\{\frac{1}{\wh{\rm{MAD}}_{\alpha}},\frac{\omega_0}{\sigma}\}$, where $\wh{\rm{MAD}}_{\alpha}=\mathrm{median}\{|y_1|,\cdots,|y_{m_{\alpha}}|\}$ and $\omega_0>0$ is any number such that $\varphi_0(\omega)>\frac{1}{2}$ for all $\omega\in[0,\omega_0]$ (which exists for any characteristic function). Afterwards, the consistent estimators $\hat{c}_\alpha=\gamma\,\hat{t}_{\mathrm{pilot}}\wh{\lVert\tilde{\mathbf{x}}\rVert}_{2,\alpha}$, $\hat{\rho}_\alpha=\frac{\sigma}{\gamma\wh{\lVert\tilde{\mathbf{x}}\rVert}_{2,\alpha}}$, and $\theta_\alpha(\hat{c}_\alpha,\hat{\rho}_\alpha)$ can be obtained \citep{l2}.

We are ready to present the main result of the study. We assume that for each $\alpha\in(0,2]\setminus\{1\}$ there exists a constant $\bar{\pi}_\alpha\in (0,1)$, such that $(m_1,m_\alpha,N)\rightarrow\infty$,
\begin{align}
\pi_\alpha:=\frac{m_\alpha}{m_1+m_\alpha}=\bar{\pi}_\alpha+o(m_\alpha^{-1/2}).\label{assu}
\end{align}
By adapting the Theorem 2 in \cite{zy} to complex-valued signals, we obtain the asymptotic property for $\widehat{k}_{\alpha,2d}(\tilde{\mathbf{x}})$ as follows.\\

\noindent
{\bf Proposition 2.} Let $\alpha\in(0,2]\setminus\{1\}$. If the conditions in Proposition 1 and assumption (\ref{assu}) hold, then as $(m_1,m_\alpha,N)\rightarrow\infty$, \begin{align}
\sqrt{\frac{m_1+m_\alpha}{\hat{w}_\alpha}}\left(\frac{\widehat{k}_{\alpha,2d}(\tilde{\mathbf{x}})}{k_{\alpha,2d}(\tilde{\mathbf{x}})}-1\right)\overset{D}\longrightarrow N(0,1),
\label{1}
\end{align}
where $\hat{w}_\alpha=\frac{\theta_\alpha(\hat{c}_\alpha,\hat{\rho}_{\alpha})}{\pi_\alpha}(\frac{1}{1-\alpha})^2+\frac{\theta_1(\hat{c}_1,\hat{\rho}_1)}{1-\pi_\alpha}(\frac{\alpha}{1-\alpha})^2$. Consequently, the asymptotic $1-\beta$ confidence interval for $k_{\alpha,2d}(\tilde{\mathbf{x}})$ is
\begin{align}
\bigg[\Big(1-\sqrt{\frac{\hat{w}_\alpha}{m_1+m_\alpha}}z_{1-\beta/2}\Big)\widehat{k}_{\alpha,2d}(\tilde{\mathbf{x}}),\Big(1+\sqrt{\frac{\hat{w}_\alpha}{m_1+m_\alpha}}z_{1-\beta/2}\Big)\widehat{k}_{\alpha,2d}(\tilde{\mathbf{x}})\bigg],
\label{2}
\end{align}
where $z_{1-\beta/2}$ is the $(1-\beta/2)$-quantile of the standard normal distribution.\\

\noindent
{\bf Remark 2.} (\ref{2}) is obtained by applying delta method to (\ref{1}) to avoid division which can cause a problem when $\Big(1-\sqrt{\frac{\hat{w}_\alpha}{m_1+m_\alpha}}z_{1-\beta/2}\Big)<0$. \\

\noindent
{\bf Remark 3.}
Since $k_{\alpha,d}(\mathbf{x})$ approaches to $\lVert \mathbf{x}\rVert_{2,0}$ as $\alpha$ approaches to 0 and $k_{\alpha,d}(\mathbf{x})=k_{\alpha,2d}(\tilde{\mathbf{x}})$, we can use $\widehat{k}_{\alpha,2d}(\tilde{\mathbf{x}})$ to approximate $\lVert \mathbf{x}\rVert_{2,0}$ with small $\alpha$.

\section{Numerical {\color{black}justification}}
In this section, we conduct some numerical experiments to confirm with the theoretical properties and show how the estimator behaves under different parameter settings. In addition, we study the sensitivity of recovery algorithm to the sparsity.
\subsection{Experimental design}

 We consider the signal $\mathbf{x}\in\mathbb{C}^N$ of the form  \begin{align}\frac{\sqrt{2}c}{2}(1+\sqrt{-1})\Big(\underbrace{\frac{1}{\sqrt{d}},\cdots,\frac{1}{\sqrt{d}}}_{d},\underbrace{\frac{1/\sqrt{d}}{2},\cdots,\frac{1/\sqrt{d}}{2}}_{d},\cdots,\underbrace{\frac{1/\sqrt{d}}{\lVert \mathbf{x}\rVert_{2,0}},\cdots,\frac{1/\sqrt{d}}{\lVert \mathbf{x}\rVert_{2,0}}}_{d},0,\cdots,0\Big) \label{sim}
 \end{align}
with $c\in\mathbb{R^{+}}$ chosen so that $\lVert \mathbf{x}\rVert_{2,2}=\lVert \mathbf{x}\rVert_2=1$ which makes it simple to calculate $k_{2,d}(\mathbf{x})$. Through the specific design of $\mathbf{x}$, the signal can be easily simulated as either strictly sparse or compressible signal via adjusting the value of $\lVert \mathbf{x}\rVert_{2,0}$. We convert the complex-valued signal $\mathbf{x}$ with block size $d$ to the real-valued signal \begin{align}
\tilde{\mathbf{x}}=\frac{\sqrt{2}c}{2}(\underbrace{\frac{1}{\sqrt{d}},\cdots,\frac{1}{\sqrt{d}}}_{2d},\underbrace{\frac{1/\sqrt{d}}{2},\cdots,\frac{1/\sqrt{d}}{2}}_{2d},\cdots,
\underbrace{\frac{1/\sqrt{d}}{\lVert \mathbf{x}\rVert_{2,0}},\cdots,\frac{1/\sqrt{d}}{\lVert \mathbf{x}\rVert_{2,0}}}_{2d},0,\cdots,0),\label{sim1}
\end{align}
with length $2N$ and block size $2d$.\\

Throughout the simulation studies, we set the signal length $N=10^3$ and the noise to be white Gaussian (i.e. $F_0$ is standard normal). Performance of the proposed estimator for block sparsity is evaluated under the following setups:
\begin{enumerate}[label=(\alph*)]
\item Verify results (\ref{1}) and (\ref{2}) with $m_1=m_2=500$ and $\sigma=0$:
\begin{itemize}
\item Sparsity estimation: we set $d=1$ and $\lVert \mathbf{x}\rVert_{2,0}=100$. We illustrate the asymptotic normality of the ratio in Proposition 2: $\frac{\widehat{k}_{2,2}(\tilde{\mathbf{x}})}{s_{2}(\mathbf{x})}$ or $\frac{\widehat{k}_{2,2}(\tilde{\mathbf{x}})}{k_{2,1}(\mathbf{x})}$. We replicate 500 times.\\

\item Block sparsity estimation: we set $d=5$ and $\lVert \mathbf{x}\rVert_{2,0}=100$. We illustrate the asymptotic normality of the ratio in Proposition 2: $\frac{\widehat{k}_{2,2d}(\tilde{\mathbf{x}})}{k_{2,d}(\mathbf{x})}$. We replicate 500 times. \\
\end{itemize}
\item Repeat (a) with $m_1=m_2=500$ and $\sigma=0.1$.
\item Repeat (a) with $m_1=m_2=1000$ and $\sigma=0$.
\item Repeat (a) with $m_1=m_2=1000$ and $\sigma=0.1$.

\item Investigate the limiting behavior of $\widehat{k}_{\alpha,2d}(\tilde{\mathbf{x}})$ with a small $\alpha$ under different settings for $\lVert \mathbf{x}\rVert_{2,0}$, i.e. $\lVert \mathbf{x}\rVert_{2,0}=10, 50, 100, 200$, respectively. Set $m_1=m_\alpha=500$, $\sigma=0$, $\alpha=0.05$ and $d=5$. We replicate 500 times.

\item Repeat (e) with $m_1=m_\alpha=1000$.
\item We set $d=1, \alpha=2$ and $\lVert \mathbf{x}\rVert_{2,0}=100.$ Investigate the sensitivity of asymptotic normality (16) to $\sigma$. Set $m_1=m_\alpha=1000$, $\sigma=0, 0.1, 0.3$ and $0.5$, respectively. We replicate 500 times.
\end{enumerate}

\subsection{Block sparse signal recovery}
Further, we investigate whether an accurate estimation of block sparsity of a block sparse signal plays an important role in the signal recovery process. To measure the block sparsity $\ell_2/\ell_0$, we use $\widehat{k}_{\alpha,2d}(\tilde{\mathbf{x}})$ with small $\alpha$ to approximate the $\lVert \mathbf{x}\rVert_{2,0}$. The model based CoSaMP algorithm proposed in \citet{Baraniuk2010} to recover an unknown signal $\mathbf{x}\in\mathbb{C}^N$ is used here. To make the recovery process fast and the signal easy to illustrate, we simulate a new complex-valued signal with a shorter length and simpler structure, i.e. $\mathbf{x}$ is of length 300 and block sparse with $\lVert \mathbf{x}\rVert_{2,0}=12$ and $d=4$, the measurement matrix $A\in\mathbb{R}^{120\times300}$ is a Gaussian random matrix, and assume that there is no measurement error. The support of $\mathbf{x}$ is a random index set and both the real and imaginary parts of its non-zero elements follow a standard Gaussian distribution. Intuitively, the best recovery should be obtained when the block sparsity is the true block sparsity. However, to our best knowledge, there is no research that shows how sensitive the recovery algorithm is to the sparsity. If the algorithm is sensitive to the input, it provides evidence and motivation to carry out this study. Relative error (RE) is used to measure the diversity between recovered signal $\mathbf{\hat{x}}$ and the true signal $\mathbf{{x}}$ by
\begin{align*}
\text{RE}=\frac{\lVert\mathbf{\hat{x}}-\mathbf{x}\rVert_2}{\lVert\mathbf{x}\rVert_2}
\end{align*}
As $\mathbf{y}$ and $A$ are random, for every block sparsity, we recover the signal $\mathbf{x}$ 100 times and take the average of REs (MRE)
\begin{align}
\text{MRE}=\frac{1}{100}\sum_{i=1}^{100}\frac{\lVert\mathbf{\hat{x}}_i-\mathbf{x}\rVert_2}{\lVert\mathbf{x}\rVert_2}
\end{align}
A plot of MREs against the different values of the block sparsity can illustrate the sensitivity of the algorithm to the block sparsity.

\section{{\color{black}Simulation results}}
\subsection{Asymptotic properties}
In this part, we show the simulation results according to the different designs mentioned above. Figure 1 illustrates the asymptotic property of $\widehat{k}_{\alpha,2d}(\tilde{\mathbf{x}})$, which corresponds to the first four designs (a)-(d) in Section 3.1. The top panel shows the behavior of $\widehat{k}_{\alpha,2d}(\tilde{\mathbf{x}})$ with $d=1, \alpha=2$ and $\lVert \mathbf{x}\rVert_{2,0}=100.$ The top left plot shows the $95\%$ confidence intervals (CIs) according to (\ref{2}) with $m_1=m_2=500, 1000$ and $\sigma=0$. Red dots and pink band represent the point estimates and CIs of $k_{\alpha,2d}(\tilde{\mathbf{x}})$ for $m_1=m_2=1000$, and the black dots and grey band represent its estimates and CIs for $m_1=m_2=500$. The black dot line is the true block sparsity of the signal. It is clear that the larger the sample size of measurements $m_1$ and $m_2$, the shorter the CIs. The same plot with $\sigma=0.1$ has the similar pattern but wider CIs (not shown). The top right one is used to verify the result (\ref{1}). The red curve is the reference, i.e. density of standard normal distribution. In practice, as $m_1,m_\alpha$ and $N$ become larger enough and $\sigma$ is reasonable small (otherwise $\theta_\alpha(c_\alpha,\rho_\alpha)$ can be non-positive),  the density of the left part of (\ref{1}) should be closer to the reference curve. In the plot, the green and purple curves represent the densities with larger sample sizes ($m_1=m_2=1000$) and $\sigma=0, 0.1,$ respectively. These two cures are closer to the red cure than the blue and yellow ones, which are densities with smaller sample sizes ($m_1=m_2=500$) and $\sigma=0, 0.1,$ respectively. The two plots from the bottom panel show the behavior of $\widehat{k}_{\alpha,2d}(\tilde{\mathbf{x}})$ with $d=5, \alpha=2$ and $\lVert \mathbf{x}\rVert_{2,0}=100$, which have the same patterns as the ones from the top panel.

\begin{figure}
    \centering
\includegraphics[scale=0.35]{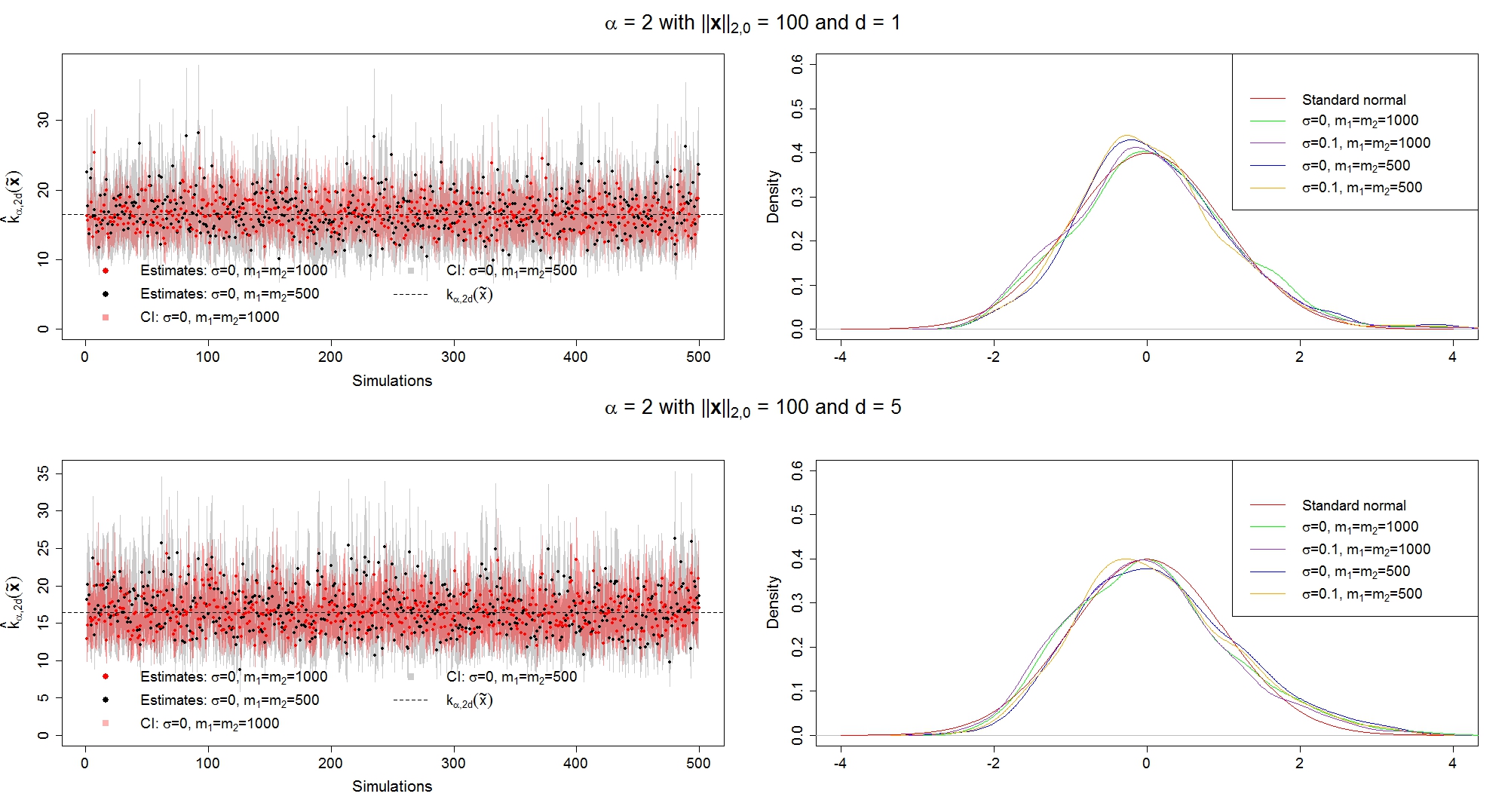}
\caption{\small {Left panel: Confidence intervals and point estimates of $k_{\alpha,2d}(\tilde{\mathbf{x}})$ under different settings over 500 simulations. Right panel: Normalized densities under different settings.}}
\end{figure}

Figure 2 is used to present the limiting behavior of $\widehat{k}_{\alpha,2d}(\tilde{\mathbf{x}})$ with a small $\alpha=0.05$ under different values for $\lVert \mathbf{x}\rVert_{2,0}$, i.e. designs (e) and (f). Since the behaviors of the estimate under $\sigma=0$ and $0.1$ are almost the same (shown in Figure 1), we only consider with $\sigma=0.1$ in this figure which is closer to reality than the noise free case. The green cures are densities with larger sample sizes ($m_1=m_2=1000$), the blue cures represent densities with smaller sample sizes ($m_1=m_2=500$). Green and blue dot lines indicate the sample means from 500 simulations with the two sample sizes, respectively. The red dot line is the true block sparsity ${k}_{\alpha,2d}(\tilde{\mathbf{x}})$ which almost overlaps with the green and blue dot lines, thus it verifies (\ref{1}) in terms of unbiasedness. Black dot lines are the true $\lVert \mathbf{x}\rVert_{2,0}$. It shows that the sample means with the two different sample sizes are very close to each other, especially, in the left bottom plot they are exactly the same, and all the sample means are close to the true $\lVert \mathbf{x}\rVert_{2,0}$ to some extent. The densities with larger sample sizes have smaller variance. Relative error in each plot measures the closeness between sample mean $\hat{\mu}$ under ($m_1=m_2=500$) and the true $\lVert \mathbf{x}\rVert_{2,0}$ through \[\frac{|\lVert \mathbf{x}\rVert_{2,0}-\hat{\mu}|}{\lVert\mathbf{x}\rVert_{2,0}}.\] It can be seen that the relative error becomes larger as $\lVert \mathbf{x}\rVert_{2,0}$ becomes larger when $\alpha$ is fixed. In practice, it guides us that smaller $\alpha$ is preferred to approximate $\lVert \mathbf{x}\rVert_{2,0}$ when a signal is less block sparse.

\begin{figure}
    \centering
\includegraphics[scale=0.35]{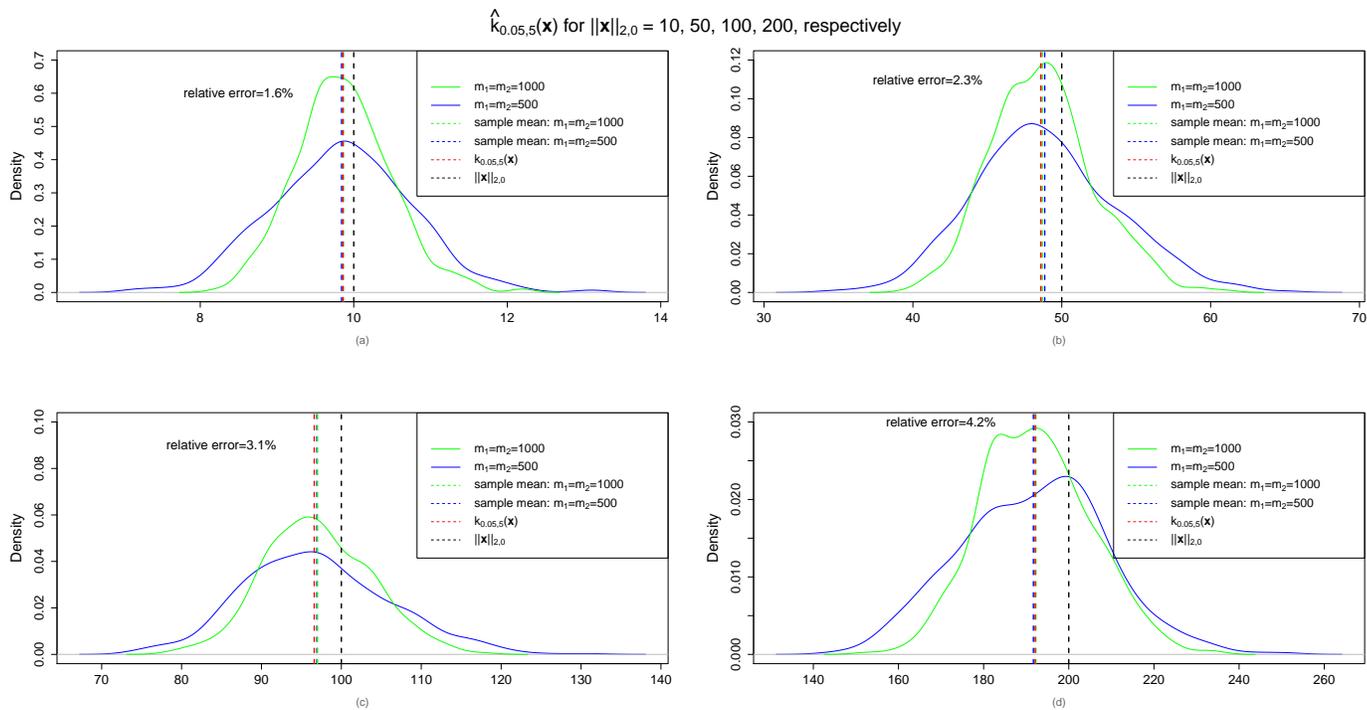}
\caption{\small {Density plots of $\widehat{k}_{0.05,10}(\tilde{\mathbf{x}})$ for $\lVert \mathbf{x}\rVert_{2,0}=10, 50, 100, 200$, respectively.}}
\end{figure}

\begin{figure}
    \centering
\includegraphics[scale=0.6]{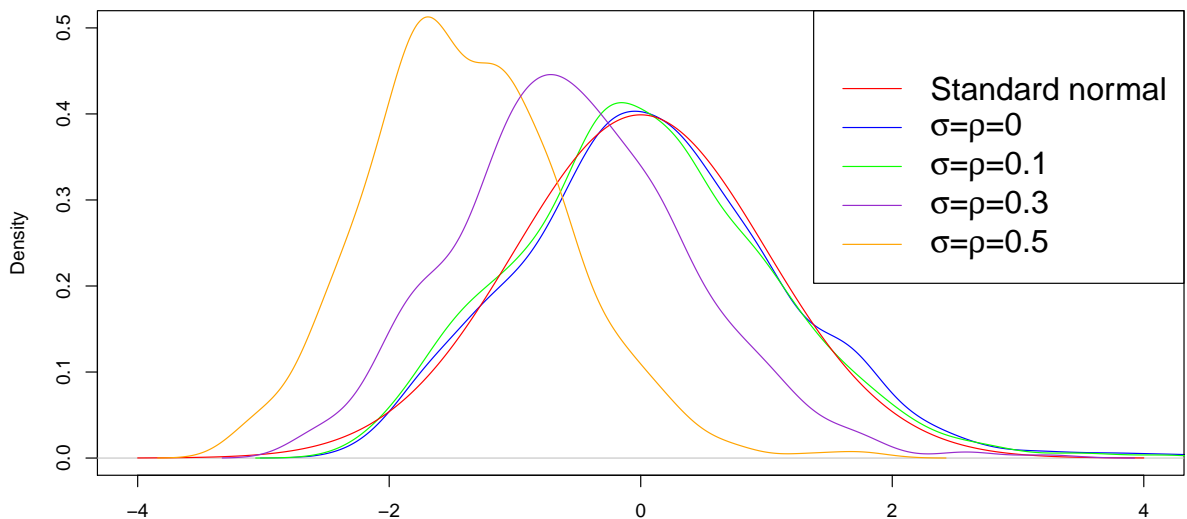}
\caption{\small {Density plots of $\sqrt{\frac{m_1+m_\alpha}{\hat{w}_\alpha}}\left(\frac{\widehat{k}_{2,2}(\tilde{\mathbf{x}})}{k_{2,2}(\tilde{\mathbf{x}})}-1\right)$ with $\lVert \mathbf{x}\rVert_{2,0}=100, m_1=m_2=1000$ and varying $\sigma$.}}
\end{figure}

Figure 3 illustrates sensitivity of the theoretical result (\ref{1}) to different $\sigma$ levels, i.e. design (g). We choose four different $\sigma=0, 0.1, 0.3$ and $0.5$, respectively. As shown in the figure, the normality approximation becomes poor as the noise increases, although Proposition 2 holds for any $\sigma\ge0$. As a result, when the noise is higher, we should increase the number of measurements $m_1$ and $m_\alpha$ as well. In this experiment, since all elements of the simulated signal (\ref{sim1}) are less than $0.3$, we should let $\sigma\le 0.1$ so that the theoretical result (\ref{1}) holds. We repeat this experiment but let $d=5$, the same conclusion is still valid, i.e. the normality approximation holds if $\sigma\le 0.1$ (not shown).

\begin{figure}
    \centering
\includegraphics[scale=0.35]{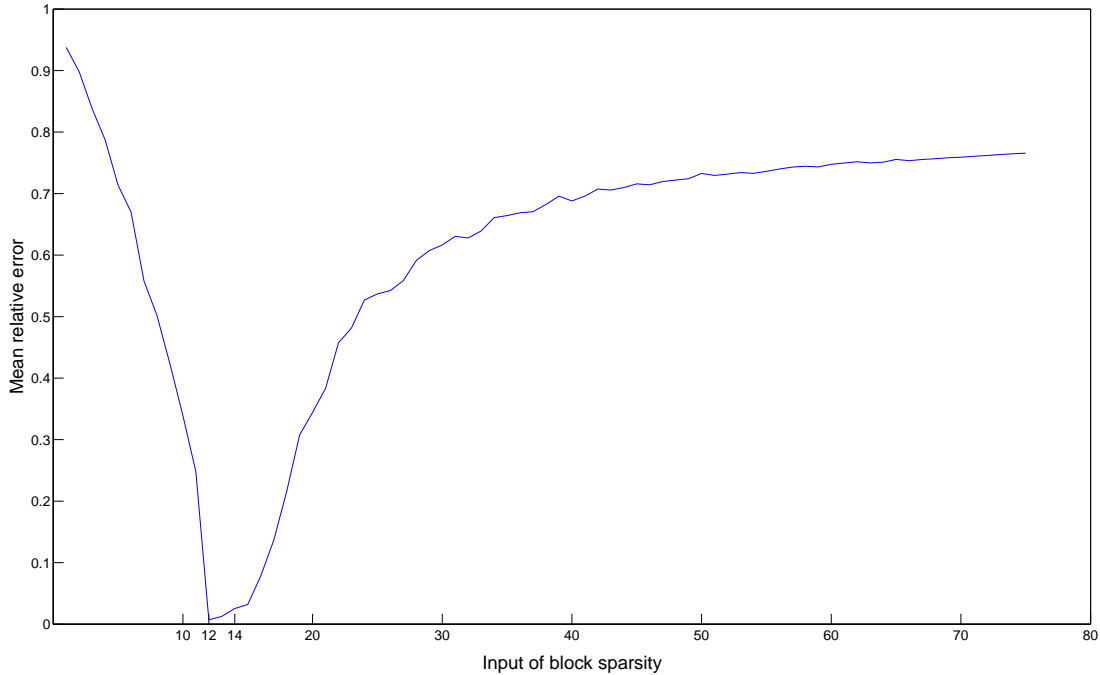}
\caption{\small {Mean relative error of recovered signals across a range of block sparsity from $0-75$.}}
\end{figure}
\subsection{Results for signal recovery}

Figure 4 shows the MRE of recovered signals over a range of block sparsity from $0{\text -}75$. Given that the number of measurements is sufficiently large, MRE reaches its minimum value when the input of block sparsity is exactly the true block sparsity of the signal, i.e. 12 in this case. The further apart from the true block sparsity, the larger MRE is. In addition, Figure 5 illustrates the true signal and two recovered signals with $\wh{||\mathbf{x}||}_{2,0} = 12$ and $7$, respectively. It is almost an exact recovery with $\wh{||{\mathbf{x}}||}_{2,0} = 12 $, however it is easy to see how diverse the recovered signal is with a less accurate block sparsity for both real and imaginary parts. Note that the sample mean of the estimates over 500 simulations with $m_1=m_2=500$ using the proposed method is $11.7$. Therefore, though this experiment we demonstrate the importance of providing an accurate estimation of block sparsity to successfully recover an unknown complex-valued block sparse signal.

\begin{figure}
    \centering
\includegraphics[scale=0.35]{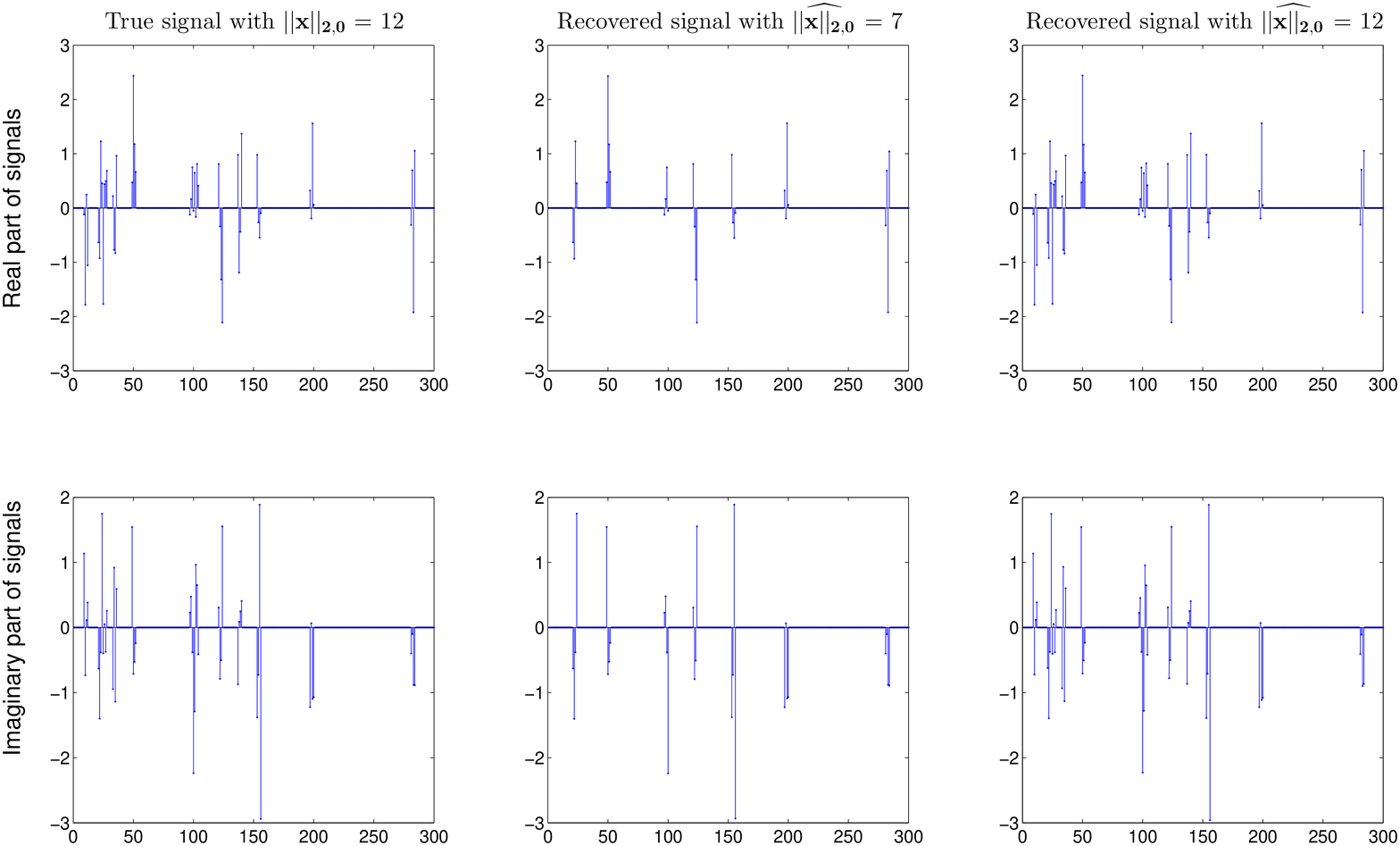}
\caption{\small {Comparison among the true signal and two recovered signals with block sparsity $\widehat{||\bf{x}||}_{2,0}$ = 12 (in the right) and 7 (in the middle).}}
\end{figure}

\section{Conclusion}
This work includes two main contributions to compressive sensing. One is that we cover up block sparsity estimation for complex-valued signals, the another is that we substantiate the importance of accurately estimating the block sparsity for signal recovery. We propose a measure of block sparsity for complex-valued signals and derive its estimator by using multivariate centered
isotropic symmetric $\alpha$-stable random projections. The measure could be used to either measure block sparsity of a block compressible signal or approximate block sparsity of a block sparse signal. The asymptotic property and limit behavior of the estimator are presented and a simulation study is conducted for verifying the theoretical results. Furthermore, the support point of our proposed estimation method is reinforced through Figure 4 and 5, which demonstrate the importance of an accurate estimation of block sparsity in the recovery algorithm.

There are still some important issues which are not covered in the study and left for further investigation. Throughout the paper, we assume that the scale parameter of noise $\sigma$ and the characteristic function of noise $\varphi_0$ are known. In practice, however, they are usually unknown and needed to be estimated. Another issue is that the measure $k_{\alpha,2d}(\tilde{\mathbf{x}})$ is valid for all $\alpha\ge 0$, while the proposed $\alpha$-stable random projection is only valid for $\alpha \in (0, 2]$. Thus a very challenging task is to find a new random projection matrix that can handle the case when $\alpha>2$. Furthermore,  how to determine the best $\alpha$ that can properly  measure the block sparsity of a signal is also an interesting and challenging problem.

\section*{Acknowledgement}
This work was supported by the Swedish Research Council grant [Reg.No. 340-2013-5342].

\bibliographystyle{plainnat}
\bibliography{sample}

\end{document}